\documentstyle[11pt]{article}

\begin{document}

\setcounter{page}{1}
\pagestyle{plain} \vspace{1cm}
\begin{center}
\Large{\bf Cosmological Inflation in a Generalized Unimodular Gravity}\\
\small \vspace{1cm} {\bf K. Nozari$^{a,b,}$\footnote{knozari@umz.ac.ir}}\quad and \quad {\bf S. Shafizadeh$^{c,}$\footnote{s.shafizadeh@tpnu.ac.ir}}\\
\vspace{0.25cm}
{\it $^{a}$ Department of Physics, Faculty of Basic Sciences,\\
University of Mazandaran,\\
P. O. Box 47416-95447, Babolsar, IRAN}\\
\vspace{0.25cm}
{\it $^{b}$ Center for Excellence in Astronomy and Astrophysics (CEAAI-RIAAM),\\
P. O. Box 55134-441, Maragha, Iran}\\
\vspace{0.5cm}
{\it $^{c}$ Department of Physics, Payame Noor University (PNU),\\
P. O. Box 19395-3697, Tehran, Iran}\\

\begin{abstract}
We study some aspects of cosmological inflation in the framework of unimodular $f(R)$ gravity. To be more clarified, we consider a generic $f(R)$ of the type $f(R)=R+\alpha R^{n}$. By
considering Einstein frame counterpart of the unimodular $f(R)$ gravity, we set the scalaron to be responsible for cosmological inflation in this setup.
We confront our model parameters space with observational data and impose some constraints on the value of $n$ in this manner. We show that for the number of e-folds $N=60$, the model is consistent with observation if $1.89<n<1.918$.
\end{abstract}
\end{center}

\vspace{1cm}
\section{Introduction}

Observational and theoretical evidences have shown that our universe has undergone two phases of positively accelerated expansion. The first phase that happened before the radiation dominated epoch in cosmic history is an early time cosmological inflation. This stage of the early universe dynamics is needed to address some problems of the standard model of cosmology. It has also the capability to explain generation of initial seeds to feed perturbations and structure formation in the universe along with temperature anisotropies observed in cosmic microwave background radiation [1]. The second phase of cosmic acceleration that happened after matter domination epoch is known as ``late time cosmic speed up". In a sense,  an unknown component dubbed ``dark energy" drives cosmic speed up at this stage [2-9]. It is well-known that standard matter with equation of state of the form $w=\frac{p}{\rho}$ cannot be responsible for driving such phases of cosmic speed up with $w\geq0$.

To provide a mechanism for cosmic speed up, the simplest candidate could be the vacuum energy density of particle physics, the cosmological constant. Although, the required cosmological constant could not absolutely alone drive the cosmic inflation, it can be considered as a candidate of dark energy to give rise late time cosmic acceleration. But cosmological constant has large energy scale to be compatible with the dark energy density [9]. An alternative proposal studied widely is called Unimodular Gravity to drop down the giant amplitude of the order of vacuum fluctuation [7,9-14]. Historically, four years after invention of general relativity, Einstein was proposed a different set of equations dubbed subsequently as \emph{unimodular gravity}. He realized also that this is equivalent to general relativity while the cosmological constant is considered as an integration constant [15-16]. In unimodular gravity one deals with those variations of the Einstein-Hilbert action that preserve the determinant of the metric to be a fixed quantity. In other words, in this scenario one assumes the determinant of the metric to be a constant as a gauge choice. An effective cosmological constant can be extracted as a constant of integration from the trace-free part of the Einstein-Hilbert field equations in unimodular gravity. This is an advantage by itself since effectively the cosmological constant is not added by hand in the theory and arises naturally as an integration constant.  

An alternative to address late time cosmic speed up is modification of the background gravitational theory. Among these theories, $f(R)$ gravities, where $R$ is the Ricci scalar, have attracted much attention at least in past two decades [17-21]. This extension has the capability to address initial inflation in a unified manner as well [22]. Modified gravity of this type replaces Ricci scalar in the Einstein-Hilbert action with a generic $f(R)$ function. In recent years a vast number of research programs have been devoted to construct $f(R)$ gravities consistent with both local (for instance Solar System) gravitational tests and cosmological observations [23-29]. These studies are extended to the issues such as spectra of galaxy clustering [30-35], cosmic microwave background radiation [24,31,36,37] and weak lensing [38,39]. The first model of modified gravity of the $f(R)$ type was introduced by Starobinsky in 1980 with $f(R)=R+\alpha R^{2}$ [22]. Starobinsky inflation model leads to accelerated expansion of the universe due to existence of $\alpha R^{2}$ term in the modified action. This model is compatible very well with the recent cosmological data [40-43]. A gravitational Lagrangian which is an arbitrary function of the Ricci scalar in the metric formalism corresponds to the generalized Brans-Dicke theory when the Brans-Dicke parameter vanishes, that is, $\omega_{BD}=0$ [44-47]. In fact, the field equations derived from $f(R)$ gravity action are the same as the equations derived from an action involving an extra degree of freedom; a scalar field [48]. This equivalence leads to a potential for a scalar field degree of freedom (that is dubbed scalaron) with a gravitational origin [22]. However, this scalar field (the scalaron) is different essentially from the scalar field in matter Lagrangian. In other words, the usual sort of the scalar field in the context of scalar-tensor theory is related to matter sector of the Lagrangian. It is relevant to energy-momentum tensor of the Einstein-Hilbert action that violates strong energy condition for inflation purpose [49,50]. This is in contrast with scalaron in $f(R)$ theory of gravity that has gravitational origin related to source of geometrical effects in Einstein-Hilbert theory. In addition, it should be pointed out that the mass of scalaron must be positive for stability of inflation theory. Moreover, in contrast to scalar field in scalar-tensor inflation models, the scalaron is massive and its rest mass depends on $R$ [51].

Recently Nojiri \emph{et al.} [52], have constructed an extension of the unimodular gravity in the spirit of general $f(R)$ theories.
They have shown that the resulting picture is different from both the ordinary $f(R)$ gravity and unimodular Einstein-Hilbert gravity.
The reason for such an extension of unimodular gravity can be explained as follows: we know that unimodular gravity is capable to address late-time acceleration via a cosmological constant which naturally arises as an integration constant. On the other hand, $f(R)$ gravity is also capable to address the cosmological constant and late time acceleration in a unified manner. A question then arises: why we need a unimodular $f(R)$ gravity? As Nojiri \emph{et al.} [52] have shown, some exotic scenarios which were not possible to realize in the standard Einstein-Hilbert gravity and also in standard unimodular gravity, now can be consistently realized in the unimodular $f(R)$ gravity theory. For instance, within the unimodular $f(R)$ gravity it is easier to implement the unified description of inflation and late time acceleration. We note also that in contrast to some strong claims were state that the cosmological constant problem can be solved in unimodular gravity, this is not actually exactly the case as has been pointed out by Padila and Saltas in Ref. [13]. In this respect, the unimodular gravity cannot be considered as the full solution of the cosmological constant problem. A unimodular $f(R)$ scenario probably has the potential to alleviate the cosmological constant problem at least to some extent. So, unimodular $f(R)$ gravity enables us to realize various cosmological scenarios which were impossible to realize in the standard Einstein-Hilbert unimodular gravity. In this respect, the authors of Ref. [52] have studied various aspects of the theory in details. They have presented also several unimodular $f(R)$ inflationary scenarios and compared their results with Planck and BICEP2 observational data. Modified unimodular gravity via Lagrange multipliers has been studied in Ref. [53]. The author has analyzed Starobinsky inflation in this framework with a comparison with its original setup. Cosmological inflation in unimodular modified teleparallel gravity is studied in Ref. [54].

In this paper we study some yet unstudied aspects of cosmological inflation in the framework of unimodular $f(R)$ gravity. Following Ref. [53], we focus on $f(R)$ theories of the type $f(R)=R+\alpha R^{n}$. We present Einstein frame counterpart of the unimodular $f(R)$ gravity with these types of modified gravity. Then we consider the scalaron to be driver of the cosmological inflation in this setup and we find slow-roll parameter in details. Finally we confront our model with observational data to find some constraints on the model parameters space. As we will show, depending on the number of e-folds parameter, one can impose severe constraint on parameter $n$. For instance, with $N=60$ the model is consistent with observation if $1.89<n<1.918$.

\section{Field Equations of Unimodular $f(R)$ Gravity }

In unimodular gravity, the cosmological constant does not appear as inserted by hand in the gravitational Einstein field equations, but it is generated as a constant of integration.
The standard way of deriving the unimodular trace-free field equations is to impose a constraint upon the determinant of the metric and setting this constraint in the action via a Lagrange multiplier. But there is another equivalent strategy too: the notion of unimodularity can be implemented also by studying the traceless part of the Einstein's field equations.  In fact, as has been shown in Ref. [53], one can depart from variational principles by using a Lagrange multiplier which imposes the unimodular constraint and leads to the trace-free part of the field equations. These trace-free part of the field equations in $n$ dimensions are as follows [52]
\begin{equation}
\label{eq1}R_{\mu\nu}-\frac{1}{n}Rg_{\mu\nu}=\kappa^{2}(T_{\mu\nu}-\frac{1}{n}Tg_{\mu\nu})\,.
\end{equation}
In this paper we consider the case $n=4$ and we set $\kappa^{2}\equiv8\pi G$ with  $T_{\mu\nu}=\frac{2}{\sqrt{-g}}\frac{\delta{\cal{L}}}{\delta g^{\mu\nu}}$.
Applying Bianchi identities shows that in contrast to the Einstein equations, the field equations (1) are not divergence-free. In fact, one finds
\begin{equation}
\label{eq2}\nabla^{\nu}R_{\mu\nu}=\frac{1}{2}\nabla_{\mu}R\,,
\end{equation}
where $R=g^{\mu\nu}R_{\mu\nu}$ is the Ricci scalar. We take divergence of the field equations (1) and then by using equation (2), we find
\begin{equation}
\label{eq3}\nabla_{\mu}R=-\frac{\kappa^{2}}{4}\nabla_{\mu}T
\end{equation}
By assuming $\nabla_{\mu}T^{\mu\nu}=0$ (that is, conservation of energy-momentum tensor) and integrating equation (3), we find
\begin{equation}
\label{eq4}R+\kappa^{2}T=4\lambda
\end{equation}
where $\lambda$ is an integration constant. Now, by inserting this result into equation (1), we find the following Einstein field equations so that $\lambda$ is introduced as an arbitrary cosmological constant [12,54]:
\begin{equation}
\label{eq5}R_{\mu\nu}-\frac{1}{2}(R+2\lambda)g_{\mu\nu}=\kappa^{2}T_{\mu\nu}\,.
\end{equation}
This is the unimodular Einstein field equation which can be obtained equivalently by
imposing a constraint upon the determinant of the metric and setting this constraint in the action via a Lagrange multiplier.
In the language of action principle, $\lambda$ here is a Lagrange multiplier in the action for unimodularity constraint on the metric determinant.
As we have explained in Introduction, unimodular $f(R)$ gravity enables us to realize various cosmological scenarios which were impossible to realize in the standard Einstein-Hilbert unimodular gravity. So it is interesting to see the status of cosmological dynamics and specially cosmological inflation in a unimodular extension of $f(R)$ gravity (see also [52,53]). A unimodular $f(R)$ gravity can be explained by a 4-dimensional action (with $f(R)$ as a generic function of the Ricci scalar) as follows [52]
\begin{equation}
\label{eq6}S=\frac{1}{2\kappa^{2}}\int d^{4}x[\sqrt{-g}(f(R)-\lambda)+\lambda] +S_{matter}\,.
\end{equation}
Variation of this action with respect to $\lambda$ gives the unimodularity constraint, $\sqrt{-g}=constant$. Also variation of this action with respect to the metric
$g_{\mu\nu}$ leads to the following field equations
\begin{equation}
\label{eq7}f_{,R}R_{\mu\nu}-\frac{1}{2}(f(R)-\lambda)g_{\mu\nu}-\nabla_{\mu}\nabla_{\nu}f_{,R}+g_{\mu\nu}\Box f_{,R}=\kappa^{2}T_{\mu\nu}^{(matter)}
\end{equation}
where $f_{,R}\equiv\frac{d f}{d R}$. Our purpose is to construct a generalization of the unimodular modified gravity of the form $f(R)=R+\alpha R^{n}$ and investigate its cosmological dynamics with focus on initial inflation. We set the gravitational action of this model in the Jordan frame as follows [52,53]
\begin{equation}
\label{eq8}S=\frac{1}{2\kappa^{2}}\int d^{4}x [\sqrt{-g}(R+\alpha R^{n})-2\lambda(\sqrt{-g}-s_{0})]+S_{matter}
\end{equation}
where $\lambda$ is a Lagrange multiplier and $s_{0}$ is a constant. Variation of this action with respect to $\lambda$ gives $\sqrt{-g}=s_{0}$.
Varying the action (8) with respect to the metric gives the following field equations
\begin{equation}
\label{eq9}(1+n\alpha R^{n-1})R_{\mu\nu}-\frac{1}{2}(R+\alpha R^{n})g_{\mu\nu}+(g_{\mu\nu}\Box-\nabla_{\mu}\nabla_{\nu})(1+n\alpha R^{n-1})+ g_{\mu\nu}\lambda=\kappa^{2}T_{\mu\nu}^{(matter)}
\end{equation}
The divergence of the field equations by using Bianchi identities shows that $\lambda$ has a constant value. The trace of equation (9) is given by
\begin{equation}
\label{eq10}(3\Box-R)(1+n\alpha R^{n-1})-2(R+\alpha R^{n})-4\lambda=\kappa^{2}T^{(matter)}
\end{equation}
where $T^{(matter)}=g^{\mu\nu}T_{\mu\nu}^{(matter)}$ and $\Box=\frac{1}{\sqrt{-g}}\partial_{\mu}(\sqrt{-g}g_{\mu\nu}\partial_{\nu})$.
The field equations in a spatially flat Friedmann-Robertson-Walker background are as [52,53]
\begin{equation}
\label{eq11}3(1+n\alpha R^{n-1})H^{2}=\frac{1}{2}(n-1)\alpha R^{n}-3n(n-1)\alpha H R^{n-2}\dot{R}+\kappa^{2}\lambda
\end{equation}
and
$$\label{eq12} -n\alpha R^{n-1}(3H^{2}+2\dot{H})\simeq$$
\begin{equation}
\Big(n(n-1)(n-2)\alpha R^{n-3}\Big)\dot{R}^{2}+2n(n-1)\alpha H\dot{R}R^{n-2}+n(n-1)\alpha R^{n-2}\ddot{R}+\frac{1}{2}\Big((1-n)\alpha R^{n}-2\lambda\Big),
\end{equation}
where to achieve cosmic acceleration we need $(1+n\alpha R^{n-1})\gg1$. By using the approximation $(1+n\alpha R^{n-1})\simeq n\alpha R^{n-1}$, we can divide the field equation by $3n\alpha R^{n-1}$ to find
\begin{equation}
\label{eq13}3H^{2}\simeq\frac{n-1}{2n}(R-6nH\frac{\dot{R}}{R})+(\frac{\kappa^{2}}{n\alpha})\lambda R^{1-n}
\end{equation}
For $n=2$, the action leads to the Starobinsky's model with $\alpha=\frac{1}{6M^{2}}$ where the WMAP normalization of the CMB temperature anisotropies constrains the mass scale to be $M\simeq10^{13}$ GeV [55].
We can use the equivalence of the scalar-tensor theory (Brans-Dicke gravity [56] with the Brans-Dicke parameter $w_{BD}=0$) and gravitational higher order theory as a scalar-tensor theory with the following action (here we use the equivalency $\kappa^{2}\equiv8\pi G$)
\begin{equation}
\label{eq14}S_{\phi}=\frac{1}{16\pi G}\int d^{4}x \bigg[\sqrt{-g}\bigg((1+n\alpha\phi^{n-1})R-(n-1)\alpha\phi^{n}\bigg)
-2\lambda(\sqrt{-g}-s_{0})\bigg]\,.
\end{equation}
Variation with respect to the scalar field gives
\begin{equation}
\label{eq15}n(n-1)\alpha\phi^{n-2}(R-\phi)=0\,,
\end{equation}
which for $\alpha\neq0$ gives $R=\phi$. For $\alpha=0$ one recovers general relativity [57]. The equation of motion shows that this equivalency leads to a potential for the scalar field degree of freedom (the scalaron, [22]) with a purely gravitational origin . Now we can easily eliminate $\phi$ by rewriting it in the form of a function of $\Phi$ with $\Phi\equiv\frac{(1+n\alpha\phi^{n-1})}{G}$ so that
\begin{equation}
\label{eq16}S_{\Phi}=\frac{1}{16\pi}\int d^{4}x\bigg[\sqrt{-g}\bigg(\Phi R-2\Lambda(\Phi)\bigg)-\frac{2}{G}\lambda(\sqrt{-g}-s_{0})\bigg]
\end{equation}
where
\begin{equation}
\label{eq17}\Lambda(\Phi)=\frac{(n-1)}{2G \alpha^{\frac{1}{n-1}}n^{\frac{n}{n-1}}}(G \Phi-1)^{\frac{n}{n-1}}\,.
\end{equation}
So, the field equations are obtained as follows
\begin{equation}
\label{eq18}R_{\mu\nu}-\frac{1}{2}g_{\mu\nu}(\Phi R-2\Lambda(\Phi))+(g_{\mu\nu}\Box-\nabla_{\mu}\nabla_{\nu})\Phi+g_{\mu\nu}\lambda=0\,.
\end{equation}
Since $\nabla_{\mu}G^{\mu\nu}=0$, then the divergence of the trace-free part of the Einstein field equations (13) gives
\begin{equation}
\label{eq19}\Phi R-4\Lambda(\Phi)-3\Box\Phi=4\lambda\,.
\end{equation}
We note that the conformal equivalence between scalar-tensor theory and general relativity with a scalar field is not just a dynamical equivalence. However, the situation for relation between scalar-tensor and higher-order gravity theories is different and the conformal equivalence in this case is a dynamical equivalence [57]. Hence, we continue our study in the Einstein frame in what follows.

\section{Field Equations in Einstein Frame}
We note that our purpose to work in Einstein frame is to analyze how a conformal transformation affects the gauge choice imposed initially, and also to see the effects of unimodular gravity in the Einstein frame. In contrast to the situation in the Jordan frame, the determinant of the metric in the Einstein frame is no longer a constant. In comparison with the situation in the Jordan frame where a cosmological constant naturally arises, here the scalar potential gets modified where results in some corrections to solutions. As show, by transforming the gravitational action from the Jordan to the Einstein frame, the determinant of the metric is not fixed to be a constant anymore. However, the Lagrange multiplier used to fix the determinant of the metric turns out to be a constant as well, such that the corresponding counterpart in the Einstein frame becomes the usual quintessence-like model but in this case with a correction in the scalar potential (see also [53] for more details). We first consider a conformal transformation to write action (16) in Einstein frame. By considering the re-scaled metric as
\begin{equation}
\label{eq120}\tilde{g}_{\mu\nu}=\Omega^{2}g_{\mu\nu}
\end{equation}
the conformal factor $\Omega^{2}=G \Phi$ is given by
\begin{equation}
\label{eq21}\psi=-\sqrt{\frac{3}{2\kappa^{2}}}\ln(G \Phi)^{-1}
\end{equation}
and therefore
\begin{equation}
\label{eq22}\tilde{g}_{\mu\nu}=e^{-\psi}g_{\mu\nu}\,.
\end{equation}
So, we obtain the action in Einstein frame as (see also [53])
\begin{equation}
\label{eq23}\tilde{S}=\int d^{4}x\bigg\{\sqrt{-\tilde{g}}\Big[\frac{1}{2\kappa^{2}}\tilde{R}
-\frac{1}{2}\tilde{g}^{\mu\nu}\partial_{\mu}\psi\partial_{\nu}\psi-V(\psi)\Big]
-2\tilde{\lambda}\Big(\sqrt{-\tilde{g}}\exp(-2\sqrt{\frac{2\kappa^{2}}{3}}\psi)-s_{0}\Big)\bigg\}\,.
\end{equation}
In this setup the potential is given by 
\begin{equation}
\label{eq24}V(\psi)=\frac{1}{2\kappa^{2}}\Bigg\{\bigg(\frac{1}{\alpha n}\bigg)^{\frac{1}{n-1}}
\bigg(1-\exp(\sqrt{(2\kappa^{2}/3)}\psi)\bigg)^{\frac{n}{n-1}}\exp\bigg(\Big(\frac{n-2}{n-1}\Big)\sqrt{(2\kappa^{2}/3)}\psi\bigg)
\bigg(\frac{n-1}{n}\bigg)\Bigg\}\,,
\end{equation}
and the determinant of the metric is given by $\sqrt{-\tilde{g}}=s_{0}\exp(2\sqrt{2\kappa^{2}/3}\psi)$.

The field equations can be deduced by varying the action in Einstein frame with respect to the conformal metric $\tilde{g}_{\mu\nu}$ (see also [53])
\begin{equation}
\label{eq25}\tilde{R}_{\mu\nu}-\frac{1}{2}g_{\mu\nu}\tilde{R}=\kappa^{2}\bigg(\partial_{\mu}\psi\partial_{\nu}\psi
-g_{\mu\nu}\Big(\frac{1}{2}\partial_{\alpha}\psi\partial^{\alpha}\psi+V(\psi)\Big)-2\tilde{\lambda}g_{\mu\nu}
\exp(-2\sqrt{2\kappa^{2}/3}\psi)\bigg).
\end{equation}
The equation of motion for the scalar field $\psi$ in conformal frame is obtained by taking variation of the action (23) with respect to this scalar field
\begin{equation}
\label{eq26}\Box\psi-V_{,\psi}(\psi)+4\tilde{\lambda}\sqrt{2\kappa^{2}/3}\exp\Big(-2\sqrt{2\kappa^{2}/3}\psi\Big)=0.
\end{equation}
The FRW equations are
\begin{equation}
\label{eq27}3\tilde{H}^{2}=\frac{1}{2}\Big(\frac{d\psi}{d\tilde{t}}\Big)^{2}+
V(\psi)+\tilde{\lambda}\exp\Big(-2\sqrt{2\kappa^{2}/3}\psi\Big)
\end{equation}
and
\begin{equation}
\label{eq28}\Big(3\tilde{H}^{2}+2\frac{d\tilde{H}}{d\tilde{t}}\Big)=
\frac{1}{2}\Big(\frac{d\psi}{d\tilde{t}}\Big)^{2}-V(\psi)
-\tilde{\lambda}\exp\Big(-2\sqrt{2\kappa^{2}/3}\psi\Big)
\end{equation}
where
\begin{equation}
\label{eq29}\tilde{H}\equiv\frac{1}{\tilde{a}}\frac{d\tilde{a}}{d\tilde{t}}=
\frac{1}{\sqrt{G\Phi}}\bigg(H+\frac{\dot{\Phi}}{2\Phi}\bigg).
\end{equation}
Also one has $V(\psi)=\frac{\Lambda(\Phi)}{(G \Phi)^{2}}$ and $\tilde{\lambda}=\frac{\lambda}{2\kappa^{2}}$.
Finally, the total potential in this setup is as follows [53]
\begin{equation}
\label{eq30}V_{tot}(\psi)=V(\psi)+2\tilde{\lambda}\Big[\exp\Big(-2\sqrt{\frac{2\kappa^{2}}{3}}\psi\Big)\Big]\,.
\end{equation}

After these preliminaries, now we study some yet unstudied features of cosmological inflation in this setup where scalaron with a purely gravitational origin is responsible for driving inflation.

\section{Slow-Roll Parameters}

We consider the slow-roll approximations as $(d\psi/d\tilde{t})^{2}\ll V_{tot}(\psi)$ and $|d^{2}\psi/d\tilde{t}^{2}|\ll|\tilde{H}d\psi/d\tilde{t}|$ during the inflation stage. By imposing these approximations to field equations in Einstein frame, we have $3\tilde{H}\simeq\kappa^{2}V_{tot}(\psi)$ and $ 3\tilde{H}(d\psi/d\tilde{t})+V_{tot,\psi}\simeq0$. Equivalently, the slow-roll parameters can be translated to the potential language as
\begin{equation}
\label{eq31}\tilde{\epsilon}\equiv-\frac{d\tilde{H}/d\tilde{t}}{\tilde{H}^{2}}\simeq
\frac{1}{2\kappa^{2}}\bigg(\frac{V_{tot,\psi}}{V_{tot}}\bigg)^{2},\,\,\,\,\,\,\,\,\
\tilde{\eta}\equiv\frac{d^{2}\psi/d\tilde{t}^{2}}{\tilde{H}(d\psi/d\tilde{t})}\simeq
\tilde{\epsilon}-\frac{1}{\kappa^{2}}\frac{V_{tot,\psi\psi}}{V_{tot}}
\end{equation}
where using (30), we obtain
\begin{equation}
\label{eq32}\tilde{\epsilon}=\frac{4}{3}\frac{\Bigg[(\frac{1}{\alpha n})^{\frac{1}{n-1}}\bigg(1-\exp(\sqrt{2\kappa^{2}/3})\bigg)^{\frac{1}{n-1}}
\exp\big(\frac{2n-3}{n-1}\sqrt{2\kappa^{2}/3}\psi
\big)+2\exp(\sqrt{4\kappa^{2}/3}\psi)V-4\lambda\Bigg]^{2}}{\Bigg[\big(\frac{1}{\alpha n}\big)^{\frac{1}{n-1}}\bigg(1-\exp(\sqrt{2\kappa^{2}/3})\bigg)^{\frac{n}{n-1}}
\exp\big(\frac{2n-3}{n-1}\sqrt{2\kappa^{2}/3}\psi
\big)(\frac{n-1}{n})+4\lambda\Bigg]^{2}}\,,
\end{equation}
and
\begin{eqnarray}
\label{eq33}\tilde{\eta}\simeq\tilde{\epsilon}-\frac{2}{3}\frac{\big(\frac{1}{\alpha n}\big)^{\frac{1}{n-1}}\exp(\sqrt{\frac{2\kappa^{2}}{3}}\psi)
\bigg[\bigg(\exp\big(-\sqrt{2\kappa^{2}/3}\psi\big)-1\bigg)^{\frac{2-n}{n-1}}
-3\bigg(\exp\big(-\sqrt{2\kappa^{2}/3}\psi\big)-1\bigg)^{\frac{1}{n-1}}\exp(\sqrt{2\kappa^{2}/3}\psi)\bigg]}
{\bigg[\big(\frac{1}{\alpha n}\big)^{\frac{1}{n-1}}\big(1-\exp(\sqrt{2\kappa^{2}/3}\psi)\big)^{\frac{n}{n-1}}
\exp(\frac{3n-4}{n-1}\sqrt{2\kappa^{2}/3}\psi)(\frac{n-1}{n})+4\lambda\bigg]}\\\nonumber
+\frac{2}{3}\frac{4V\exp(\sqrt{\frac{2\kappa^{2}}{3}}\psi)+\lambda}{\bigg[\big(\frac{1}{\alpha n}\big)^{\frac{1}{n-1}}\big(1-\exp(\sqrt{2\kappa^{2}/3}\psi)\big)^{\frac{n}{n-1}}
\exp(\frac{3n-4}{n-1}\sqrt{2\kappa^{2}/3}\psi)(\frac{n-1}{n})+4\lambda\bigg]}\,,
\end{eqnarray}
respectively. If we set $\lambda=0$ (that is, without unimodular gravity), we recover corresponding equations for inflation in $f(R)$ modified gravity. Specially, for $n=2$ we recover the standard Starobinsky inflation. When $\alpha\lambda^{n-1}>\exp[(4-3n)\sqrt{2\kappa^{2}/3}\psi]$, to have enough number of e-folds before the end of inflation, one finds
\begin{equation}
\label{eq34}\tilde{\epsilon}=\frac{1}{3}\bigg[4+\frac{n}{(n-1)}\Big(1-\exp(\sqrt{2\kappa^{2}/3}\psi)\Big)^{-2}
-\frac{4n}{(n-1)}\Big(1-\exp(\sqrt{2\kappa^{2}/3}\psi)\Big)^{-1}\bigg]\,,
\end{equation}
and
\begin{equation}
\label{eq35}\tilde{\eta}\simeq\tilde{\epsilon}-\frac{8}{3}-
\frac{n}{(1-n)}\bigg(2\Big[1-\exp(\sqrt{2\kappa^{2}/3}\psi)\Big]^{-1}
\bigg)\,.
\end{equation}
With these equations we are in the position to confront this model with observational data. 

\section{Parameters Space Analysis and Confrontation with Observation}

Now we study slow-roll inflation parameters numerically in this unimodular modified gravity. We do this end in two steps: first we consider the case that unimodularity is absent ($\lambda=0$) or has so small effect relative to modified $f(R)$ gravity for different values of $n$.  For $\lambda=0$, or for  $(M^{2}/\lambda)^{n-1}>\exp((4-3n)\sqrt{2\kappa^{2}/3}\psi)$ with $\alpha=1/(6M^{2})^{n-1}$ we focus on the scalar field potential for different values of $n$ and we compare our numerical results with Planck2015 joint observational data [44].
The Starobinsky model, corresponding to $n=2$, has been studied in [22,55]. The inflationary expansion of the universe in this model provided by the higher derivative terms probably comes from supergravity [58,59] in the Lagrangian which takes the form ${\cal{L}}=R+\alpha R^{2}$. A simple numerical analysis for Starobinsky inflation in this model gives
$\tilde{n}_{s}=0.9654$ and $\tilde{r}=0.003$ which are well in the range of observational data.
To proceed further, we confront the model with observational data in this unimodular modified gravity for $n<2$. To have accelerated expansion we need $n>\frac{5}{4}$.
For $\frac{5}{4}<n<2$, accelerating expansion happens due to falling the scalar field from the value of $\psi=0$ toward the minimum of the potential at $\psi=-\infty$ in Einstein frame. In this case, the scalar field potential (30) takes the following form
\begin{equation}
\label{eq57}V(\psi)=\frac{1}{2\kappa^{2}}\Bigg\{\bigg(\frac{1}{\alpha n}\bigg)^{\frac{1}{n-1}}
\exp\bigg(\frac{(n-2)}{n-1}\sqrt{(2\kappa^{2}/3)}\psi\bigg)
\bigg(\frac{n-1}{n}\bigg)\Bigg\}\,.
\end{equation}
So, the number of e-folds is given by
\begin{equation}
\label{eq58}N=\int^{\psi_{end}}_{\psi_{i}}\Big(-\frac{\tilde{H}}{\dot{\psi}}\Big)d\psi\approx
-\frac{\sqrt{6}}{2}\frac{(n-1)}{n-2}\kappa\Big(\psi_{end}-\psi_{i}\Big)\,.
\end{equation}
Then we can derive the initial scalar field versus e-folds number as
\begin{equation}
\label{eq59}\psi_{i}\approx\frac{\sqrt{6}}{3\kappa}\frac{(n-2)N}{(n-1)}\,.
\end{equation}
The slow roll parameters are derived as
\begin{equation}
\label{eq60}\tilde{\epsilon}\simeq\frac{1}{3}\frac{(n-2)^{2}}{(n-1)^{2}} \,,\quad\quad
\mid\tilde{\eta}\mid\simeq\frac{4}{3}\frac{(2-n)}{(n-1)}\,.
\end{equation}
The spectral index in this model can be expressed as
\begin{equation}
\label{eq61}\tilde{n}_{s}\simeq1-\frac{8}{3}\frac{(2-n)}{(n-1)}\,,\quad\quad
r\simeq\frac{16}{3}\frac{(n-2)^{2}}{(n-1)^2}\,,
\end{equation}
and finally the power spectrum of primordial scalar perturbation in Einstein frame is given by
\begin{equation}
\label{eq62}\tilde{{\cal{P}}}_{\Psi}=\frac{1}{24\pi^{2}}\frac{\Big(\frac{1}{\alpha n}\Big)^{\frac{1}{n-1}}
\exp\Big(\frac{\sqrt{6}(n-2)\psi}{3(n-1)}\Big)(n-1)^{3}}{n(n-2)^{2}}\,.
\end{equation}
Figure 1 shows the tensor-to-scalar ratio versus the spectral index in this setup in the background of Planck2015 joint data when unimodularity is ignored (that is, $\lambda=0$ or $(M^{2}/\lambda)^{n-1}>\exp((4-3n)\sqrt{2\kappa^{2}/3}\psi)$ with $\alpha=1/(6M^{2})^{n-1}$). The Starobinsky inflation is shown with point $n=2$ in this figure. As the figure shows for  $1.888<n<1.915$ there is good agreement between this model and Planck2015 TT, TE, EE+lowP Joint data.

\begin{figure}
\begin{center}\includegraphics{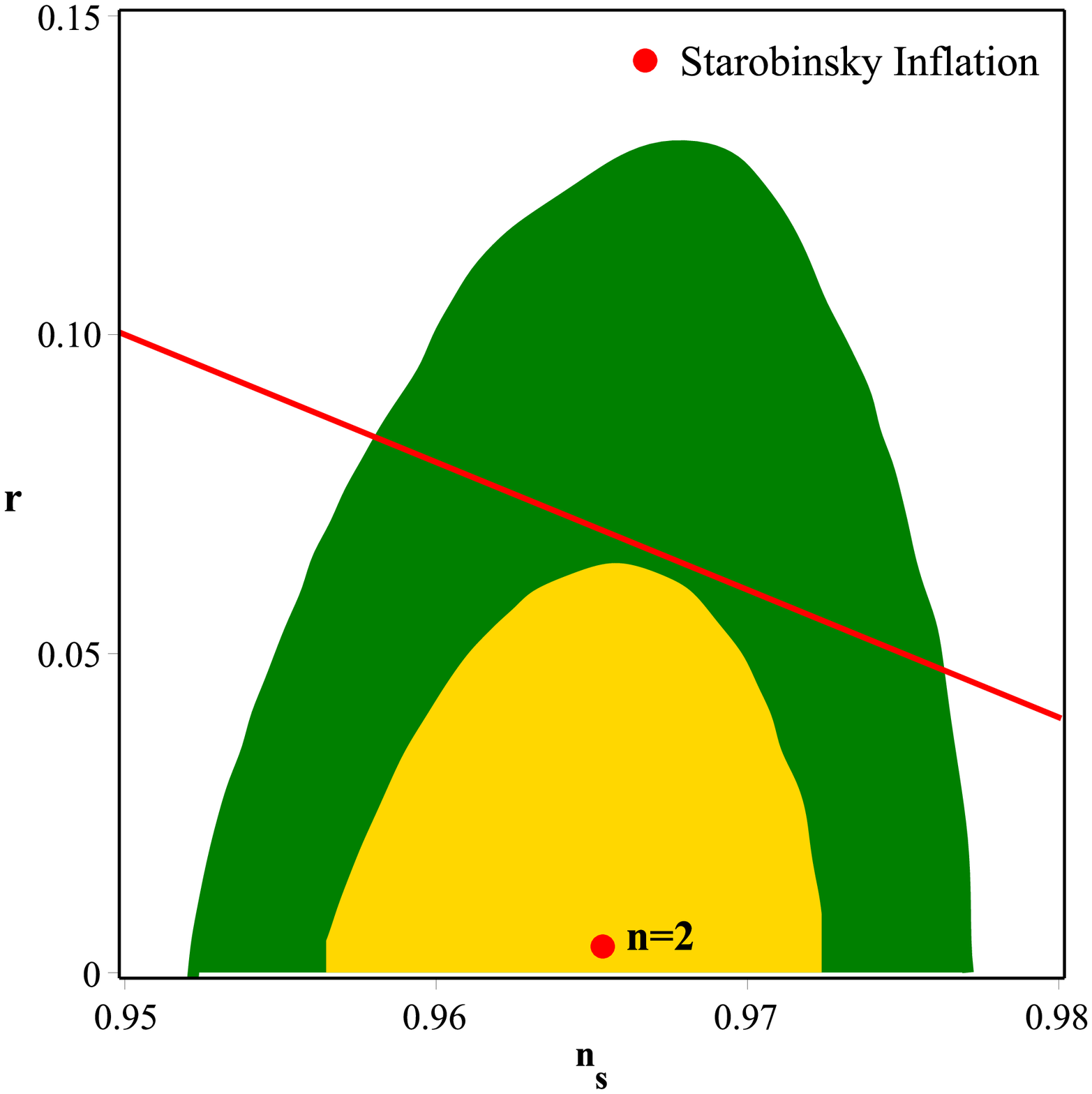} \vspace{7cm}
\end{center}
\caption{\small {Tensor-to-scalar ratio versus the scalar spectral index without unimodularity in the background of Planck2015 TT, TE, EE+lowP data.}}
\end{figure}

Now we consider the case with $n>2$ in the action (14) with $\lambda=0$ (or $(M^{2}/\lambda)^{n-1}>\exp((4-3n)\sqrt{2\kappa^{2}/3}\psi)$). In the Jordan frame by using equations (11)-(13) we can obtain $\dot{H}$ as follows
\begin{equation}
\label{eq63}\dot{H}\simeq-\frac{(n-1)(4-2n)}{6n(n+4(n-1)^{2})}\bigg[R-\frac{6nH\dot{R}}{R}\bigg].
\end{equation}
The Hubble parameter grows up with time in this case which leads to high curvature limit and there is no possibility for exit from the inflation phase if $n>2$ (see also [14,55]). Nevertheless, in this case we have the de Sitter phase for $\dot{H}=0$ that can be analyzed in Einstein frame. To impose the de Sitter condition to equation (13) (that can be obtained from equation of motion (15) with $R=\phi$) and also by using equation (21) we find
\begin{equation}
\label{eq64}\psi=-\sqrt{\frac{3}{2\kappa^{2}}}\ln\bigg(\frac{2(n-1)}{n-2}\bigg)\,.
\end{equation}
This solution corresponds to the potential of the scalar field in Einstein frame as follows
\begin{equation}
\label{eq65}V(\psi)=(\alpha n)^{\frac{1}{1-n}}\bigg(\frac{1}{n-2}\bigg)^{\frac{n-2}{1-n}}\,.
\end{equation}
Therefore, the slow-roll parameters for this case are given as follows 
\begin{equation}
\label{eq66}\tilde{\epsilon}\simeq0 \,,\quad\quad\mid\tilde{\eta}\mid\simeq\frac{(2-n)}{6(n-1)}(n)^{\frac{2-n}{n-1}}\,.
\end{equation}
The spectral index for the de Sitter solution is given by
\begin{equation}
\label{eq67}\tilde{n}_{s}\simeq1-\frac{(n-2)}{6(n-1)}(n)^{\frac{2-n}{n-1}}\,.
\end{equation}
As we have mentioned, since the potential of the scalar field is constant during the inflation in this case, there is no possibility of graceful exit from inflation phase in this case.

In the final case we focus mainly on the effect of unimodularity on inflation parameters in this setup. In this case, we remember that for $n<2$ the Hubble parameter is redefined as follows \begin{equation}
\label{eq68}\tilde{H}^{2}=\frac{1}{6\kappa^{2}}\Bigg\{\bigg(\frac{1}{\alpha n}\bigg)^{\frac{1}{n-1}}
\exp\bigg(\frac{(n-2)}{n-1}\sqrt{(2\kappa^{2}/3)}\psi\bigg)
\bigg(\frac{n-1}{n}\bigg)+\lambda\Big(\exp(-2\sqrt{2\kappa^{2}/3}\psi)\Big)\Bigg\}\,,
\end{equation}
and the equation of motion of the field during the inflation stage reads
\begin{eqnarray}
\label{eq69}3\tilde{H}\dot{\psi}\simeq-\frac{1}{2\kappa^{2}}\Bigg[\bigg(\frac{1}{\alpha n}\bigg)^{\frac{1}{n-1}}
\exp\bigg(\frac{(n-2)}{n-1}\sqrt{(2\kappa^{2}/3)}\psi\bigg)
\frac{(n-2)}{n}\sqrt{2\kappa^{2}/3}
\\+\bigg(-2\lambda\sqrt{2\kappa^{2}/3}\bigg)\exp\bigg(-2\sqrt{2\kappa^{2}/3}\psi\bigg)\Bigg]\,.\nonumber
\end{eqnarray}
The number of e-folds in this unimodular $f(R)$ inflation is given by
\begin{equation}
\label{eq70}N=\sqrt{6}\psi\bigg(\frac{1-n}{2n-4}+4\bigg)+\frac{3}{4(n-2)}
\ln\Biggl(\exp\bigg(\frac{(n-2)\sqrt{6}\psi}{3(n-1)}\bigg)-\frac{2\lambda\exp\big(-2/3 \sqrt{6}\psi\big)n}{(\alpha n)^{1/1-n}(n-2)}\Biggl)\big(1-n\big)
\end{equation}
and the slow-roll parameters are obtained as
\begin{equation}
\label{eq71}\tilde{\epsilon}=\frac{1}{2}\frac{\Biggl[\frac{1}{\sqrt{6}}
\frac{(n-2)}{n}\Big(\frac{1}{\alpha n}\Big)^{\frac{1}{n-1}}
\exp\Bigg(\frac{4}{3}\frac{N(n-2)^{2}}{n(n-1)}\Bigg)-\frac{2}{3}\lambda \sqrt{6}
\exp\Bigg(-\frac{8}{3}\frac{N(n-1)}{n}\Bigg)\Biggl]^{2}}{\Biggl[\frac{1}{2}
\frac{(n-2)}{n}\Big(\frac{1}{\alpha n}\Big)^{\frac{1}{n-1}}
\exp\Bigg(\frac{4}{3}\frac{N(n-2)^{2}}{n(n-1)}\Bigg)+\frac{1}{2}\lambda
\exp\Bigg(-\frac{8}{3}\frac{N(n-2)}{n}\Bigg)\Biggl]^{2}}
\end{equation}
and
\begin{equation}
\label{eq72}\tilde{\eta}=\tilde{\epsilon}-\frac{\Biggl[\frac{1}{3}
\frac{(n-2)^{2}}{n(n-1)}\Big(\frac{1}{\alpha n}\Big)^{\frac{1}{n-1}}
\exp\Bigg(\frac{4}{3}\frac{N(n-2)^{2}}{n(n-1)}\Bigg)-\frac{2}{3}\lambda \sqrt{6}
\exp\Bigg(-\frac{8}{3}\frac{N(n-1)}{n}\Bigg)\Biggl]}{\Biggl[\frac{1}{2}
\frac{(n-2)}{n}\Big(\frac{1}{\alpha n}\Big)^{\frac{1}{n-1}}
\exp\Bigg(\frac{4}{3}\frac{N(n-2)^{2}}{n(n-1)}\Bigg)+\frac{1}{2}\lambda
\exp\Bigg(-\frac{8}{3}\frac{N(n-2)}{n}\Bigg)\Biggl]}
\end{equation}
respectively. Figure 2 shows the case that unimodularity is included. In this figure the tensor-to scalar ratio is plotted versus the spectral index for three different values of the number of e-folds. For $N=50$ the model is well in the confidence levels of Planck2015 TT, TE, EE+lowP data if $1.9065<n<1.922$. For $N=55$ we find $1.902<n<1.920$ and for $N=60$ the model is consistent with data if $1.89<n<1.918$.

\begin{figure}
\begin{center}\includegraphics{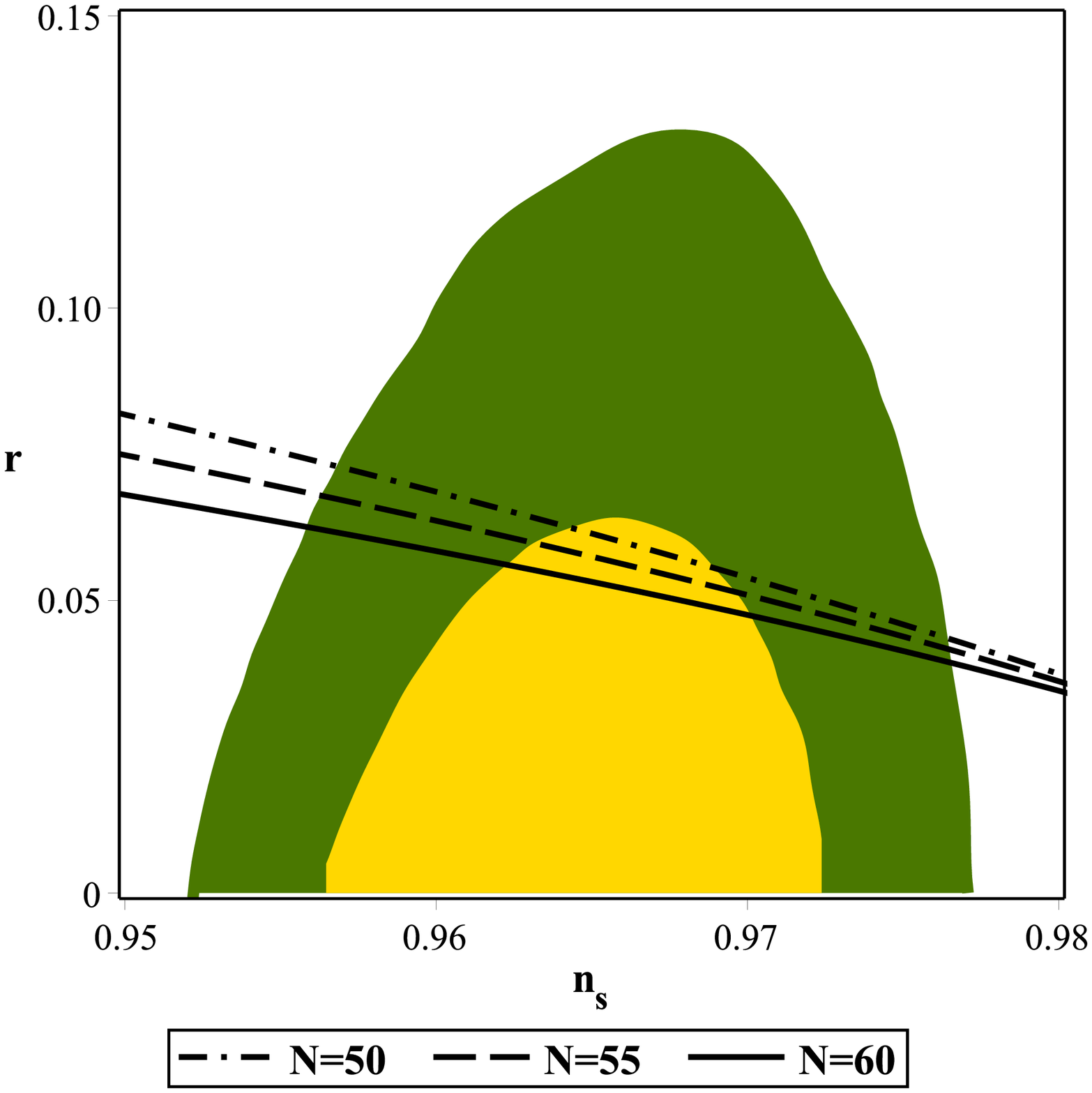} \vspace{7cm}
\end{center}
\caption{\small {Tensor-to-scalar ratio versus the scalar spectral index in unimodular modified gravity in the background of Planck2015 TT, TE, EE+lowP data.}}
\end{figure}

\section{Summary and Conclusions}

In this paper we have studied cosmological inflation in an extension of $f(R)$ modified gravity in the spirit of unimodular gravitational scenarios.
Unimodular gravity introduces cosmological constant naturally in a fascinating manner and this feature makes its possible extensions interesting in cosmological setups. 
The reason for such an extension is that some exotic scenarios which are not possible to be realized in the standard Einstein-Hilbert gravity and also in standard unimodular gravity, now can be consistently realized in the unimodular $f(R)$ gravity theory. For instance, within the unimodular $f(R)$ gravity it is easier to implement a unified description of inflation and late time acceleration. While the issue of cosmological inflation in $f(R)$ unimodular gravity has been studied in Refs. [52,53], here we focused on cosmological viability of unimodular $f(R)$ theories with $f(R)=R+\alpha R^{n}$ in confrontation with observational data. We have presented Einstein frame counterpart of the unimodular $f(R)$ gravity with the mentioned types of modified gravity and then we have set the scalaron to be responsible for cosmological inflation in this setup. We studied slow-roll inflation in this framework and the results are compared numerically with observational data from Planck2015 TT, TE, EE+lowP data. In this manner we were able to see the status of unimodular inflation which is consistent with observation. For the chosen form of $f(R)$ function we have obtained severe constraints on $n$ for viability of the cosmological inflation in confrontation with observation for different numbers of e-folds. With $N=60$ the model is well in the confidence levels of Planck2015 TT, TE, EE+lowP data if $1.89<n<1.918$. Also for $N=50$ and $N=55$ this model is consistent with observation if $1.9065<n<1.922$ and $1.902<n<1.920$ respectively.

{\bf Acknowledgement}\\
The work of K. Nozari has been supported financially by Center for Excellence in
Astronomy and Astrophysics of IRAN (CEAAI-RIAAM) under research project No. 1/4717-120. 
We are grateful to a referee for very insightful comments that improved the work considerably.

\end{document}